\let\includefigures=\iftrue
%
\let\useblackboard=\iftrue
%
%
\newfam\black
\input harvmac
\noblackbox
\includefigures
\message{If you do not have epsf.tex (to include figures),}
\message{change the option at the top of the tex file.}
\input epsf
\def\figin{\epsfcheck\figin}\def\figins{\epsfcheck\figins}
\def\epsfcheck{\ifx\epsfbox\UnDeFiNeD
\message{(NO epsf.tex, FIGURES WILL BE IGNORED)}
\gdef\figin##1{\vskip2in}\gdef\figins##1{\hskip.5in}
\else\message{(FIGURES WILL BE INCLUDED)}%
\gdef\figin##1{##1}\gdef\figins##1{##1}\fi}
\def\DefWarn#1{}
\def\figinsert{\goodbreak\midinsert}
\def\ifig#1#2#3{\DefWarn#1\xdef#1{fig.~\the\figno}
\writedef{#1\leftbracket fig.\noexpand~\the\figno}%
\figinsert\figin{\centerline{#3}}\medskip\centerline{\vbox{
\baselineskip12pt\advance\hsize by -1truein
\noindent\footnotefont{\bf Fig.~\the\figno:} #2}}
\bigskip\endinsert\global\advance\figno by1}
\else
\def\ifig#1#2#3{\xdef#1{fig.~\the\figno}
\writedef{#1\leftbracket fig.\noexpand~\the\figno}%
\global\advance\figno by1}
\fi
%

\def\smallfig#1#2#3{\DefWarn#1\xdef#1{fig.~\the\figno}
\writedef{#1\leftbracket fig.\noexpand~\the\figno}%
\figinsert\figin{\centerline{#3}}\medskip\centerline{\vbox{
\baselineskip12pt\advance\hsize by -1truein
\noindent\footnotefont{\bf Fig.~\the\figno:} #2}}
\endinsert\global\advance\figno by1}

\useblackboard
\message{If you do not have msbm (blackboard bold) fonts,}
\message{change the option at the top of the tex file.}
\font\blackboard=msbm10 scaled \magstep1
\font\blackboards=msbm7
\font\blackboardss=msbm5
\textfont\black=\blackboard
\scriptfont\black=\blackboards
\scriptscriptfont\black=\blackboardss

\else

\fi
%
\def\yboxit#1#2{\vbox{\hrule height #1 \hbox{\vrule width #1
\vbox{#2}\vrule width #1 }\hrule height #1 }}
\def\fillbox#1{\hbox to #1{\vbox to #1{\vfil}\hfil}}
\def\ybox{{\lower 1.3pt \yboxit{0.4pt}{\fillbox{8pt}}\hskip-0.2pt}}
%
%

\def\comments#1{}



\def\II{\relax{I\kern-.10em I}}

\def\IZ{\relax\ifmmode\mathchoice
{\hbox{\cmss Z\kern-.4em Z}}{\hbox{\cmss Z\kern-.4em Z}}
{\lower.9pt\hbox{\cmsss Z\kern-.4em Z}}
{\lower1.2pt\hbox{\cmsss Z\kern-.4em Z}}
\else{\cmss Z\kern-.4emZ}\fi}
\def\IB{\relax{\rm I\kern-.18em B}}
\def\IC{{\relax\hbox{$\inbar\kern-.3em{\rm C}$}}}
\def\ID{\relax{\rm I\kern-.18em D}}
\def\IE{\relax{\rm I\kern-.18em E}}
\def\IF{\relax{\rm I\kern-.18em F}}
\def\IG{\relax\hbox{$\inbar\kern-.3em{\rm G}$}}
\def\IGa{\relax\hbox{${\rm I}\kern-.18em\Gamma$}}
\def\IH{\relax{\rm I\kern-.18em H}}
\def\II{\relax{\rm I\kern-.18em I}}
\def\IK{\relax{\rm I\kern-.18em K}}
\def\IP{\relax{\rm I\kern-.18em P}}

%

\def\inbar{\,\vrule height1.5ex width.4pt depth0pt}

\font\cmss=cmss10 
\def\IR{\relax{\rm I\kern-.18em R}}

%

\def\BP{\IP}

\def\lp10{\ell_p^{10}}
\def\lp11{\ell_p^{11}}
\def\R11{R_{11}}

\def\frac#1#2{{#1 \over #2}}


\hyphenation{Di-men-sion-al}



\lref\Myers{R. Myers, ``New Dimensions for Old Strings,'' Phys. Lett.
{\bf B199} (1987) 371.} 

\lref\Zurab{Z. Kakushadze, ``Self-tuning and Conformality,'' hep-th/0009199.}

\lref\HW{P. Horava and E. Witten, ``Heterotic and Type I String
Dynamics from Eleven Dimensions,'' Nucl. Phys. {\bf B460} (1996)
506, hep-th/9510209.}

\lref\candelas{P. Candelas, M. Lynker and R. Schimmrigk, ``Calabi-Yau
Manifolds in Weighted P(4),'' Nucl. Phys. {\bf B341} (1990) 383.} 

\lref\kklmI{S. Kachru, S. Katz, A. Lawrence and J. McGreevy,
``Open String Instantons and Superpotentials,'' Phys. Rev.
{\bf D62} (2000) 026001, hep-th/9912151.}

\lref\kklmII{S. Kachru, S. Katz, A. Lawrence and J. McGreevy,
``Mirror Symmetry for Open Strings,'' hep-th/0006047.} 

\lref\bdlr{I. Brunner, M. Douglas, A. Lawrence and C. Romelsberger,
``D-branes on the Quintic,'' JHEP {\bf 0008} (2000) 015, hep-th/9906200.}

\lref\syz{A. Strominger, S.T. Yau and E. Zaslow, ``Mirror Symmetry
is T Duality,'' Nucl. Phys. {\bf B479} (1996) 243, hep-th/9606040.}

\lref\Dave{D. Morrison, ``Geometric Aspects of Mirror Symmetry,''
math.ag/0007090.}

\lref\McGreevy{S. Kachru and J. McGreevy, ``Supersymmetric Three Cycles
and Supersymmetry Breaking,'' Phys. Rev. {\bf D61} (2000) 026001,
hep-th/9908135.}

\lref\Witten{E. Witten, ``The Cosmological Constant from the Viewpoint of
String Theory,'' hep-ph/0002297.} 

\lref\Douglas{M. Douglas, ``Topics in D-Geometry,'' Class. Quant. Grav.
{\bf 17} (2000) 1057, hep-th/9910170.}

\lref\cdgp{P. Candelas, X. de la Ossa, P. Green and
L. Parkes, ``A Pair of Calabi-Yau Manifolds as an Exactly Soluble
Superconformal Theory,'' Nucl. Phys. {\bf B359} (1991) 21.}

\lref\Katz{D. Cox and S. Katz, {\it Mirror Symmetry and Algebraic Geometry},
Mathematical Surveys and Monographs No. 68, American Mathematical Society, 1999.}

\lref\Greene{B. Greene, ``String Theory on Calabi-Yau Manifolds,''
hep-th/9702155.} 

\lref\cdo{P. Candelas and X. de la Ossa, ``Moduli Space of 
Calabi-Yau Manifolds,'' Nucl. Phys. {\bf B355} (1991) 455.}

\lref\Joe{J. Polchinski, ``Dirichlet Branes and Ramond-Ramond Charges,''
Phys. Rev. Lett. {\bf 75} (1995) 4724, hep-th/9510017.}

\lref\ooy{H. Ooguri, Y. Oz and Z. Yin, ``D-Branes on Calabi-Yau Spaces
and their Mirrors,'' Nucl. Phys. {\bf B477} (1996) 407, hep-th/9606112.}

\lref\Frank{J. Feng, J. March-Russell, S. Sethi and F. Wilczek, ``Saltatory
Relaxation of the Cosmological Constant,'' hep-th/0005276.}

\lref\MikeTom{T. Banks, M. Dine and L. Motl, ``On Anthropic Solutions of the
Cosmological Constant Problem,'' hep-th/0007206.}

\lref\McLean{R. McLean, ``Deformations of Calibrated Submanifolds,''
Duke University PhD thesis, Duke preprint 96-01: see www.math.duke.edu/
preprints/1996.html.}

\lref\Nilles{S. Forste, Z. Lalak, S. Lavignac and H. Nilles, ``The Cosmological
Constant Problem from a Brane World Perspective,'' JHEP {\bf 0009} (2000) 034,
hep-th/0006139.}

\lref\Ed{E. Witten, ``Bound States of Strings and P-branes,'' Nucl. Phys.
{\bf B460} (1996) 335, hep-th/9510135.}

\lref\RSI{L. Randall and R. Sundrum, ``A Large Mass Hierarchy from
a Small Extra Dimension,'' Phys. Rev. Lett. {\bf 83} (1999) 3370,
hep-th/9905221.}

\lref\RSII{L. Randall and R. Sundrum, ``An Alternative to 
Compactification,'' Phys. Rev. Lett. {\bf 83} (1999) 4690, hep-th/9906064.}
 
\lref\AdS{J. Maldacena, ``The Large N Limit of Superconformal
Field Theories and Supergravity,'' Adv. Theor. Math. Phys.
{\bf 2} (1998) 231, hep-th/9711200\semi
S. Gubser, I. Klebanov and A. Polyakov, ``Gauge Theory Correlators from
Noncritical String Theory,'' hep-th/9802109\semi 
E. Witten, ``Anti-de Sitter Space and Holography,'' Adv. Theor. Math. Phys.
{\bf 2} (1998) 253, hep-th/9802150.}

\lref\WW{X.G. Wen and E. Witten, ``World Sheet Instantons and
the Peccei-Quinn Symmetry,'' Phys. Lett. {\bf B166} (1986) 397.} 

\lref\DSWW{M. Dine, N. Seiberg, X.G. Wen and E. Witten, ``Nonperturbative
Effects on the String World Sheet,'' Nucl. Phys. {\bf B278} (1986)
769.}

\lref\sw{N. Seiberg and E. Witten, ``Electric-Magnetic Duality,
Monopole Condensation, and Confinement in ${\cal N}=2$ Supersymmetric
Yang-Mills Theory,'' Nucl. Phys. {\bf B426} (1994) 19, hep-th/9407087.} 

\lref\Gubser{J. Maldacena, unpublished\semi
E. Witten, unpublished\semi
S. Gubser, ``AdS/CFT and Gravity,'' hep-th/9912001.}

\lref\Giddings{S. Giddings, E. Katz and L. Randall, ``Linearized
Gravity in Brane Backgrounds,'' JHEP {\bf 0003} (2000) 023, hep-th/0002091.}

\lref\KKS{S. Kachru, J. Kumar and E. Silverstein, ``Vacuum Energy
Cancellation in a Nonsupersymmetric String,'' Phys. Rev. {\bf D59}
(1999) 106004, hep-th/9807076.}

\lref\Tom{T. Banks, ``Cosmological Breaking of Supersymmetry?  Or
Little Lambda Goes Back to the Future 2,'' hep-th/0007146.} 

\lref\Herman{H. Verlinde, ``Holography and Compactification,''
Nucl. Phys. {\bf B580} (2000) 264, hep-th/9906182.}

\lref\GSW{M. Green, J. Schwarz and E. Witten, ``Superstring Theory:
Volume II,'' Cambridge University Press, 1987.}

\lref\Wise{W. Goldberger and M. Wise, ``Modulus Stabilization with
Bulk Fields,'' Phys. Rev. Lett. {\bf 83} (1999) 4922, hep-th/9907447.}

\lref\FV{P. Frampton and C. Vafa, ``Conformal Approach to
Particle Phenomenology,'' hep-th/9903226.}

\lref\KS{S. Kachru and E. Silverstein, ``4d Conformal Field Theories
and Strings on Orbifolds,'' Phys. Rev. Lett. {\bf 80} (1998) 4855,
hep-th/9802183.}

\lref\LED{N. Arkani-Hamed, S. Dimopoulos and G. Dvali, ``The
Hierarchy Problem and New Dimensions at a Millimeter,''
Phys. Lett. {\bf B429} (1998) 263, hep-th/9803315.}

\lref\KSS{S. Kachru, M. Schulz and E. Silverstein, ``Self-tuning
Flat Domain Walls in 5d Gravity and String Theory,'' Phys. Rev.
{\bf D62} (2000) 045021, hep-th/0001206\semi
S. Kachru, M. Schulz and E. Silverstein, ``Bounds on Curved
Domain Walls in 5d Gravity,'' Phys. Rev. {\bf D62} (2000) 085003,
hep-th/0002121.}

\lref\ADKS{N. Arkani-Hamed, S. Dimopoulos, N. Kaloper and
R. Sundrum, ``A Small Cosmological Constant from a Large Extra
Dimension,'' Phys. Lett. {\bf B480} (2000) 193, hep-th/0001197.}

\lref\Verlinde{E. Verlinde and H. Verlinde, ``RG Flow, Gravity
and the Cosmological Constant,'' JHEP {\bf 0005} (2000) 034, 
hep-th/9912018.}

\lref\KSSei{S. Kachru, N. Seiberg and E. Silverstein, ``SUSY Gauge
Dynamics and Singularities of 4d ${\cal N}=1$ String Vacua,''
Nucl. Phys. {\bf B480} (1996) 170, hep-th/9605036\semi
S. Kachru and E. Silverstein, ``Singularities, Gauge Dynamics, and
Nonperturbative Superpotentials in String Theory,'' Nucl. Phys.
{\bf B482} (1996) 92, hep-th/9608194.} 

\lref\Vafamir{C. Vafa, ``Extending Mirror Conjecture to Calabi-Yau with
Bundles,'' hep-th/9804131.}

\lref\OgVafa{H. Ooguri and C. Vafa, ``Knot Invariants and Topological
Strings,'' Nucl. Phys. {\bf B577} (2000) 419, hep-th/9912123.}

\lref\Rubakov{V. Rubakov and M. Shaposhnikov, ``Extra Space-Time
Dimensions: Towards a Solution to the Cosmological Constant Problem,''
Phys. Lett. {\bf B125} (1983) 139.}

\lref\RubakovII{V. Rubakov and M. Shaposhnikov, ``Do We Live Inside
a Domain Wall?,'' Phys. Lett. {\bf B125} (1983) 136.}

\lref\BP{R. Bousso and J. Polchinski, ``Quantization of Four-Form
Fluxes and Dynamical Neutralization of the Cosmological Constant,''
JHEP {\bf 0006} (2000) 006, hep-th/0004134.}

\lref\PS{J. Polchinski and M. Strassler, ``The String Dual of a
Confining Four-Dimensional Gauge Theory,'' hep-th/0003136.}

\lref\TEVbrane{I. Klebanov and M. Strassler, ``Supergravity and
a Confining Gauge Theory: Duality Cascades and $\chi{SB}$
Resolution of Naked Singularities,'' JHEP {\bf 0008} (2000) 052,
hep-th/0007191\semi
J. Maldacena and C. Nunez, ``Towards the Large N Limit of Pure
${\cal N}=1$ Super Yang-Mills,'' hep-th/0008001.} 

\lref\Horowitz{G. Horowitz, I. Low and A. Zee, ``Selftuning in an
Outgoing Brane Wave Model,'' hep-th/0004206.}

\lref\BT{J. Brown and C. Teitelboim, ``Dynamical Neutralization of
the Cosmological Constant,'' Phys. Lett. {\bf B195} (1987) 177\semi
J. Brown and C. Teitelboim, ``Neutralization of the Cosmological
Constant by Membrane Creation,'' Nucl. Phys. {\bf B297} (1988) 787.}

\lref\Quevedo{G. Aldazabal, L. Ibanez, F. Quevedo and A. Uranga,
``D-branes at Singularities:  A Bottom-up Approach to the String
Embedding of the Standard Model,'' JHEP {\bf 0008} (2000) 002,
hep-th/0005067.}

\Title{\vbox{\baselineskip12pt\hbox{hep-th/0009247}
\hbox{SU-ITP-00/23}
\hbox{SLAC-PUB-8641}}}
{\vbox{
\centerline{Lectures on Warped Compactifications}
\medskip
\centerline{and}
\medskip
\centerline{ Stringy Brane
Constructions}}}

\smallskip
\centerline{Shamit Kachru} 
\bigskip 
\centerline{Department of Physics and SLAC}
\centerline{Stanford University}
\centerline{Stanford, CA 94305/94309}
\bigskip
\bigskip
\noindent
In these lectures, two different aspects of 
brane world scenarios in 5d gravity or string theory are 
discussed.  In the
first two lectures,  work on 
how warped compactifications of 5d gravity theories can change the
guise of the gauge hierarchy problem and the cosmological constant
problem is reviewed, and a discussion of several issues which remain
unclear in this context is provided.  In the next two lectures,
microscopic constructions in string theory which involve D-branes
wrapped on cycles of Calabi-Yau manifolds are described.  The focus
is on computing the superpotential in the brane worldvolume field
theory.  Such calculations may be a necessary step towards 
understanding e.g.
supersymmetry breaking and moduli stabilization in stringy realizations
of such scenarios, and are of intrinsic interest as probes of the
quantum geometry of the Calabi-Yau space.

\Date{September 2000}

\newsec{Introduction}

Scenarios for an underlying string theory description of nature have been
considerably enriched following the duality revolution of the mid 1990s.
Perhaps the most striking qualitative new feature is the emergence of
scenarios in which standard model gauge fields are confined to some 
submanifold of a larger bulk spacetime, while of course gravity propagates
in the bulk.  For instance, such models are natural in 
the Horava-Witten extension of the $E_8
\times E_8$ heterotic string
theory, where finite string coupling opens up an additional
dimension with the geometry of an interval, and 
the $E_8 \times E_8$ gauge fields live on the boundaries
\HW.  
More generally, after the realization of the important role played
by D-branes in string duality \Joe, it was found that the world-volume
quantum field theory on coincident D-branes enjoys a non-Abelian gauge
symmetry \Ed.  This makes it natural to construct type II or type I
string models where the standard model gauge fields are confined to
stacks of D-branes (see e.g. \Quevedo\ for a discussion of some
such attempts). 

String constructions of this sort have also motivated new ideas in
long wavelength effective field theory for reformulating the
gauge hierarchy problem \refs{\LED,\RSI} and the cosmological constant
problem \refs{\Verlinde,\KSS,\ADKS} in terms of brane world constructions.
The translation of these problems to brane world language does not
solve them, but certainly provides a different way of thinking about
them, and opens up exciting new possibilities for phenomenology. 

In the following lectures, we will first review some of the new ideas
for reformulating the hierarchy problems of fundamental physics in
the language appropriate to such ``brane world'' scenarios.  We will
then switch tracks and talk about the detailed investigation of one
class of microscopic brane constructions that exist in string theory.
These latter lectures will start with a telegraphic review of some aspects of
closed string Calabi-Yau compactifications.   They will then  
focus 
on superpotential computations in models with
4d ${\cal N}=1$ supersymmetry, since these are quite relevant
to issues of physical interest like supersymmetry breaking and stabilization
of moduli.

\newsec{Trapped Gravity and the Gauge Hierarchy}   

\subsec{Trapping Gravity}

Our world might actually be contained on a 3+1
dimensional defect, e.g. a domain wall, in a higher dimensional
spacetime \RubakovII. 
Why would we see 4d gravity in such a model? 

Suppose the extra spatial dimension, parametrized by $x_5$, is a 
circle of radius $r$, and we are localized around some point on this
circle.  
The global 5d metric looks like the metric on $R^{3,1}$ times a circle.
The 5d Einstein action is:

\eqn\fivedac{S_5 ~=~\int~d^{5}x ~\sqrt{-G} ~R~M_{5}^3}
where $M_5$ is the 5d Planck scale.  Integrating out the ``extra''
$x_5$ dimension gives rise to a 4d effective action with an
effective Planck scale
\eqn\match{M_4^2 \sim r ~M_{5}^3}

Hence, at length scales larger than $r$, gravity will appear to be
four-dimensional with a Newton's constant determined by \match.
For sufficiently small $r$, this of course would reproduce everything
we know about gravity from present day experiments.

However, there is a more general possibility.  The metric can be
$\it warped$. 
For instance, consider pure 5d gravity with a cosmological constant,
and a source term for a domain wall located at $x_5 = 0$:

\eqn\newac{S~=~\int ~d^{5}x ~\sqrt{-G}\left( R - \Lambda\right)~
+ \int~d^{4}x~\sqrt{-g} (-V_{brane})}
where $g_{\mu\nu} = \delta^M_\mu \delta^N_\nu G_{MN}(x_5 = 0)$,
$\mu,\nu = 1,\cdots,4$, and $M,N = 1,\cdots 5$.   
Following Randall and Sundrum \RSII, 
we will find solutions of \newac\
which give rise to 4d gravity and in which non-trivial warping plays an
essential role.
If we want the 4d world to look flat, we should look for solutions of
the equantions of motion following from \newac\ which exhibit an
$SO(3,1)$ symmetry (the Poincare invariance of our world).  The
most general such ansatz for the 5d metric is:

\eqn\ansatz{ds^2 = e^{2A(x_5)}\eta_{\mu\nu}dx^\mu dx^\nu + dx_5^2}

With this ansatz, Einstein's equations just become equations for the
warp factor $A$ (primes denote differentiation with respect to
$x_5$):

\eqn\eqnone{6 (A^\prime)^2 + {1\over 2}\Lambda ~=~ 0}
\eqn\eqntwo{3A^{\prime\prime} + {1\over 2}V \delta(x_5) ~=~0}
Choosing $\Lambda < 0$, a negative 5d cosmological constant, one 
can quickly solve \eqnone\ to find
\eqn\abeh{A = \pm k x_5,~~k = \sqrt{-{\Lambda\over 12}}}
Integrating \eqntwo\ from 
$x_5 = -\epsilon$ to $x_5 = \epsilon$ to pick up the delta function
contribution, one finds:
\eqn\disc{3 \Delta (A^\prime) = -{1\over 2}V}
where $\Delta$ denotes the discontinuity across $x_5 = 0$.

So to solve the equations with the ansatz \ansatz, we must take
$A = -kx_5$ for $x_5 > 0$, $A = kx_5$ for $x_5 < 0$.  
Furthermore, we must tune the brane tension $V$ in terms of the
bulk cosmological constant $\Lambda$ so that
\eqn\tune{V = 12k = 12\sqrt{-{\Lambda\over 12}}}

This yields a solution where
\eqn\adsmet{ds^2 = e^{-2k\vert x_5\vert} \eta_{\mu\nu}dx^{\mu}dx^{\nu}
+ dx_5^2}
The warp factor is sharply peaked at $x_5 = 0$, where the domain wall,
which we will call the ``Planck brane,'' is located.
This fact leads to the existence of localized gravity at the
Planck brane \RSII.  Namely, doing the naive 5d to 4d reduction by
simply integrating over the $x_5$ direction, one finds

\eqn\integ{M_4^2 = M_5^3 \int dx_5 e^{-2k\vert x_5\vert} < \infty} 
This is $\it finite$ despite the existence of an infinite 5th dimension,
so the 4d Newton's constant on the Planck brane is finite, and an
observer there would see effective four-dimensional gravity.

There is a natural concern that arises in this case, that does
not arise in the case of a 5d theory compactified on a circle of
radius $r$.  In the latter case, the lightest 5d Kaluza-Klein (KK)
modes have masses which go like ${1\over r}$.  For $r$ small enough
to avoid experimental detection, this leads to a gap in the
KK spectrum, and the low energy theory is clearly just 4d general
relativity coupled to the brane worldvolume fields.

In the warped case, however, the infinite extent of the $x_5$ dimension
means that there is $\it no ~gap$ in the spectrum of bulk modes!  So,
one should seriously worry that they will appear as particles with
a continuum of masses in 4d, and ruin 4d effective field theory.
It has been argued in e.g. \RSII\ and \Giddings\ that despite the
gapless KK spectrum, a physicist on the Planck brane would still see
effectively four-dimensional physics.  This is because, although the
KK spectrum is gapless, most of the bulk KK modes have wavefunctions
with support far from $x_5 = 0$ where the brane fields are localized.
Due to this very small overlap of wavefunctions in the $x_5$ direction,
the brane fields and localized graviton couple only very weakly to
the bulk continuum.  So for instance the Newtonian form of the 
gravitational potential $V(r) \sim {1\over r}$ receives only
small ${1\over r^3}$ corrections (curiously, as if there were two extra 
flat dimensions) \refs{\RSII,\Giddings}. 

\subsec{Hierarchies from Multiple Branes}

Consider now a case with two branes, located at $x_5 = 0$ and
$x_5 = \pi$.  We will take $x_5$ to live on the interval between
$0$ and $\pi$ , so the
space now has an extra dimension of finite extent. 
The total action looks like:
\eqn\totalac{S = \int d^5x ~\sqrt{-G} (R - \Lambda) + S_{SM} + S_{Pl}}
where $S_{SM}$ is the action on the ``Standard Model brane'' located at
$x_5 = \pi$ and $S_{pl}$ is the action on the ``Planck brane'' located
at $x_5 = 0$, i.e. 
\eqn\ssm{S_{SM} = \int d^4x \sqrt{-g_{SM}}( {\cal L}_{SM} - V_{SM})} 
\eqn\spl{S_{pl} = \int d^4x \sqrt{-g_{pl}} ( {\cal L}_{pl} - V_{pl})}
Following Randall and Sundrum \RSI, we will 
demonstrate solutions of the theory
\totalac\ which give rise to large hierarchies between scales in a somewhat
natural way.

We again look for a warped metric which maintains the 4d Poincare
invariance we desire:
\eqn\ansatz{ds^2 = e^{-2A(x_5)} \eta_{\mu\nu}dx^\mu dx^\nu + r^2 dx_5^2}
It follows from \ansatz\ that the size of the $x_5$ interval is
$\pi r$.

The Einstein equations are now:

\eqn\einsone{6 {{(A^\prime)^2} \over r^2} + {1\over 2} \Lambda = 0}
\eqn\einstwo{3 {{A^\prime\prime} \over r^2} + {1\over 2} {V_{pl}\over r}
\delta (x_5) + {1\over 2} {V_{SM}\over r} \delta (x_5 - \pi) ~=~0}   

As before, we can define $k = \sqrt{-{\Lambda \over 12}}$ (and take
the bulk $\Lambda$ to be negative).   Then, \einsone\ is solved by
taking
\eqn\soleo{A(x_5) ~=~k r \vert x_5 \vert} 
Notice that \soleo\ is consistent with a $Z_2$ symmetry under which
$x_5$ is reflected; our strategy will be to find a $Z_2$ symmetric
solution where $-\pi < x_5 < \pi$, and then orbifold by the $Z_2$
to get the desired setup.

From \soleo\ (and the periodicity of $x_5$, where $x_5 = \pm \pi$ are
identified), it follows that
\eqn\secder{A^{\prime\prime} ~=~2kr \left( \delta(x_5) - \delta(x_5 - \pi)
\right)}
Comparing this to \einstwo, we see that we should choose
\eqn\tune{V_{pl} ~=~-V_{SM} ~=~12k}
in order to find a Poincare invariant solution.

Taking this as our background gravity solution, what is the 4d
effective field theory on the 
``Standard Model'' brane that follows from it?  First of all notice
that it is natural to write the metric ansatz in terms of 
4d fields as follows (here $x$ runs over the dimensions other than
$x_5$):

\eqn\fourdf{ds^2 = e^{-2kT(x)}\left( \eta_{\mu\nu} + h_{\mu\nu}(x)\right)
dx^\mu dx^\nu + T(x)^2 dx_5^2}  
There are two dynamical 4d fields in \fourdf; the four-dimensional metric
${\overline{g}}_{\mu\nu}(x) = 
h_{\mu\nu}(x) + \eta_{\mu\nu}$, and 
the 4d scalar field $T(x)$ (the so-called ``radion'').  We recover the
desired vacuum solution by choosing the radion to have a constant VEV
$\langle T(x) \rangle = r$. 

One can easily compute the 4d effective action for the metric
${\overline g}$ by starting from the 5d Einstein action; one finds
a 4d Einstein term with
\eqn\match{M_{4}^2 ~=~M_5^3 ( 1- e^{-2kr\pi} )}
for the 4d Planck scale.  
In particular, notice that for $r$ of reasonable magnitude (in Planck
units), $M_4$ depends only very weakly on $r$.  This is intuitively
because the 4d graviton is largely localized in the vicinity of the
Planck brane, which is at $x_5 = 0.$

This leads to an interesting phenomenon.  If we compute the metrics
$g_{pl}$ and $g_{SM}$ which appear in the source terms for the Planck
and Standard Model branes, it follows that
\eqn\plme{g^{pl}_{\mu\nu} = {\overline{g}}_{\mu\nu}}
while
\eqn\smme{g^{SM}_{\mu\nu} = e^{-2kr\pi} {\overline {g}}_{\mu\nu}}

This reflects the fact that an object with energy $E$ at the Planck
brane would be seen, at the Standard Model brane, as an object with
energy $E e^{-kr\pi}$; equivalently, length scales at the Standard Model
brane are ``redshifted'' to be longer than the corresponding lengths
at the Planck brane.  This is a familiar manifestation of scale/radius
duality in AdS/CFT \AdS.  However, here it has the interesting
consequence that if one starts with dimensionful parameters in the
Standard Model Lagrangian ${\cal L}_{SM}$ (e.g. a Higgs mass) of
order the 4d Planck scale, they can easily be ``redshifted'' down
to energies which are hierarchically smaller, simply by the 
factors of the metric \smme.  Hence, to find TeV
scale physics on the Standard Model brane, one simply needs to choose
$r$ to be of order ten times the fundamental scale.  This 
sounds relatively natural, and provides a candidate solution to the
hierarchy problem.  

Finally, one should ask, how easy is it to accomplish the
stabilization of the radion around
the required value (that leads to an ``explanation'' of the hierarchy)?
Goldberger and Wise have argued that the presence of a bulk scalar
field, with fairly natural bulk and brane couplings, can do the job
\Wise.
 
\subsec{Some Remarks on Randall-Sundrum scenarios}

In this section, we make some remarks which have relevance to both
the naturalness of RS scenarios, and their possible embedding into
a fully consistent microscopic theory of gravity (like string theory).
There has been a great deal of research on this topic. 

There are several different things to say about this.
One is that, via the AdS/CFT correspondence \AdS, the Randall-Sundrum
scenario is more or less a strong coupling version of an older idea
for solving the hierarchy problem just within quantum field theory.
It has long been realized that if one starts with some ultraviolet
fixed point CFT around the UV scale (say, just below the Planck scale)
and perturbs it by a marginally relevant operator (whose dimension is
very close to $4$, say $4-\epsilon$) then one can naturally generate scales
much lower than $M_{pl}$.  Namely, the RG running of the couplings
in the perturbed field theory is logarithmic, and therefore the 
relevant coupling will have significant dynamical effects only after
a vast amount of RG running (in energy scale space).  
Roughly speaking, the scale at which the operator produces significant
dynamical effects might be $M = e^{-1/\epsilon} M_{pl}$. 
A scenario of this sort was advocated recently by
Frampton and Vafa \FV.

Obviously, for $\epsilon$ finely tuned close enough to zero, one can
achieve $M << M_{pl}$.  However, it might be quite a challenge to find
a 4d conformal field theory whose most relevant perturbation is of dimension
$4 - \epsilon$ with $\epsilon$ small; and if one cannot find such a theory,
then this mechanism becomes unnatural (because the other, more relevant
operators will also be activated along the RG flow).

This concern 
has a direct translation into the Randall-Sundrum scenario, 
via the AdS/CFT dictionary \Gubser.\foot{This clear relationship
and the corresponding concerns 
were stressed to the author by Maldacena and Witten on separate occasions.}
Their
scenario requires the existence (between the Planck and the Standard
Model branes) of a region many AdS radii in size where the
5d metric is approximately that of AdS space.
By introducing a Planck brane, they have also rendered normalizable
those modes in AdS space which are normally non-normalizable (due to
divergent behavior near the boundary of AdS).
These modes will fluctuate.  It is difficult to think of concrete
scenarios where none of these fluctuating modes in the gravity
theory is a ``tachyon'' (which maps, via the AdS/CFT duality, to a 
relevant operator).  
Such tachyons, when they fluctuate and condense, will destroy the
AdS form of the metric between the Planck and Standard Model branes.
The question of why tachyons (except perhaps those which are very
close to being non-tachyonic) should be absent, is the same as the
question raised above in the field theory picture.
This is not surprising; by AdS/CFT duality, the Randall-Sundrum
scenario is exactly the same as the previous one, except in the
limit of strong field theory 't Hooft coupling. 
Of course, such a limit was never tractable before, so interesting
new features could emerge there. 

Without addressing these concerns, one can still ask whether one
can plausibly realize the mechanism of \S2.2\ in some class of string theory
backgrounds.  A good argument that this is possible has been provided
by H. Verlinde \Herman. 
He recalls that in certain 
compactifications of F-theory on a Calabi-Yau fourfold
$X_4$, one can introduce 
\eqn\ndthree{N ~=~{{\chi(X_4)}\over 24}}
space filling D3 branes to satisfy tadpole cancellation conditions (at least if
the sign of the Euler characteristic is correct).
Since the known list of Calabi-Yau fourfolds includes some with
$\vert \chi \vert \sim 200,000$, this can lead to the introduction of
large numbers of D3 branes. 
Of course, with 4d ${\cal N}=1$ supersymmetry, dynamics will undoubtedly
dictate the positions of these D3 branes in the end (there will be a
superpotential for the chiral fields which fixes their positions).  
But it is quite plausible that large numbers of D3 branes will be
stacked on top of one another, generating an AdS throat which is
``glued'' into the CY fourfold asymptotically.  Then, the D3 brane
field theory will hopefully, in the infrared, manufacture some analogue
of the RS Standard Model brane, while the gluing of the throat to the
CY fourfold acts effectively as a Planck brane.  More explicit models
of what the TeV brane might look like have emerged in the recent papers
\TEVbrane. 

\newsec{Brane Worlds and the Cosmological Constant}

\subsec{The Problem}

It is an old idea, going back at least to Rubakov and Shaposhnikov
\Rubakov, that if the Standard Model were confined to a defect in a
higher dimensional space (e.g. a domain wall), this defect
might naturally like
to be flat.  Suitably interpreted, the flatness of the defect could
then explain the extreme smallness of the cosmological constant in
our 4d world. 

In this section, we discuss the extent to which this idea seems realizable
in the ``wall world'' scenarios which have become common today.  
To see that the idea doesn't always fare well, lets begin by reviewing
the relevant portions of the Randall-Sundrum scenario.
For simplicity, we discuss the scenario of \S2.1, but that of \S2.2 would
differ in no essential way.

So, suppose we did live on a domain wall in a 5d gravity theory with
bulk $\Lambda < 0$.  The key point is to recall that, in searching
for a Poincare invariant 4d world, we were forced by the Einstein
equations to tune the tension of the brane $V$ in terms of the 
bulk cosmological term, as expressed in equation  
\tune:
$$V ~=~12 \sqrt{-{\Lambda\over 12}}$$

Now, if we imagine the Standard Model degrees of freedom living on the
wall at $x_5 = 0$, small changes in the Standard Model parameters
(the electron mass, QCD scale, weak scale,...) will result in a
renormalization of the brane tension $V \to V + \Delta V$.  
Equivalently, quantum loops of Standard Model fields enter in $V$.
But under such a shift, the relation \tune\ will be violated, and
hence one will no longer be able to find a flat solution! 

This is the manifestation of the cosmological constant problem in 
such wall world scenarios.  One must tune the brane tension $V$, 
which depends in a sensitive way on the Standard Model parameters,
in terms of other microscopic parameters, or one cannot find a 
Poincare invariant 4d world.

\subsec{Adding Scalars}

In most microscopic theories which could be responsible for the 5d
bulk action in a wall world scenario, there are degrees of freedom
other than the 5d metric.  For instance, in string theory generic
compactifications result in massless scalar moduli.  So, it is natural
to consider a theory with additional 5d bulk scalars, and see if the
situation of \S3.1\ improves.  In fact, as discussed in \refs{\KSS,\ADKS},
it does to some extent. 

So, take now for the action:
\eqn\fivedac{S ~=~\int d^5{x} \sqrt{-G} \left(R - {4\over 3} (\nabla \phi)^2\right) 
 + \int d^4 x \sqrt{-g}\left(-f(\phi)\right)} 
In addition to the action for the 5d gravity and scalar field, we have
a source term for a domain wall at $x_5 = 0.$ 
In the presence of the scalar $\phi$, it is natural to take the wall
tension to be $\phi$ dependent.  
For instance, the tension of branes in string theory can depend on the
string dilaton, or the moduli controlling the volumes of 
cycles which they are wrapping.
Also, we have chosen to start with
an action with no 5d cosmological term.  Our philosophy throughout this
section will be that the 5d bulk is supersymmetric, while the theory on
the 4d domain wall breaks SUSY.  Hence, it is natural (in a controlled
expansion in small parameters, which we will discuss later) 
to choose the bulk to have vanishing
cosmological term. 

For simplicity, we will for now take 
\eqn\choice{f(\phi) = V e^{b\phi}}
Most of what we say generalizes to far more generic $f(\phi)$, as 
detailed in \KSS.
To look for Poincare invariant 4d worlds, we again choose the metric
ansatz:
\eqn\metric{ds^2 = e^{2A(x_5)} \eta_{\mu\nu}dx^\mu dx^\nu + dx_5^2}
and we take the scalar $\phi = \phi(x_5)$.

The resulting equations are (where again $^\prime$ denotes
differentiation with respect to $x_5$):

\eqn\emone{{8\over 3}\phi^{\prime\prime} + {32\over 3} A^\prime
\phi^\prime = bV e^{b\phi} \delta(x_5)}
\eqn\emtwo{6 (A^\prime)^2 - {2\over 3} (\phi^\prime)^2 = 0}
\eqn\emthree{3 A^{\prime\prime} + {4\over 3}(\phi^\prime)^2 = 
- {1\over 2}e^{b\phi} V \delta(x_5)}

An important fact which is immediately evident from the equations
above is that finding flat solutions will NOT require any fine
tune of the coefficient $V$
in \choice\ in terms of any microscopic parameters.  
This is obvious because the only non-derivative coupling of the
scalar $\phi$ is through the brane tension term (in $f(\phi)$). 
So given a solution for one value of $V$, a shift of $V$ to
$V + \Delta V$ can be compensated by an appropriate shift in
the zero mode of $\phi$, leaving the equations of motion unchanged.

Why is this significant?  The Standard Model physics at $x_5 = 0$
is purely reflected (in this approximation, where the theory is in
its ground state) through the wall source term.  Now, suppose the
Standard Model gauge couplings are independent of $\phi$.  
Then varying Standard Model parameters, or summing Standard Model
radiative corrections, will shift $V$ in a way that is $\phi$
independent.  Hence, one can effectively absorb any cosmological
constant generated by Standard Model physics, while still finding
a Poincare invariant 4d world \refs{\KSS,\ADKS}. 

A picture where $\phi$ is, at leading order, unrelated to 
Standard Model couplings is not unreasonable.
For instance
if we are in string theory, 
we could let the brane at $x_5 = 0$ be a D-brane and $\phi$ be a geometrical
modulus for a cycle the brane does not wrap.  Alternatively,   
if we wish to treat $\phi$ as the
dilaton, we could imagine the brane at $x_5 = 0$
is a wrapped 
NS brane whose effective gauge coupling is determined by some geometrical
modulus, as in the examples of \KSSei.  

\subsec{What about 4d gravity?}

To proceed, lets write down the explicit solutions to the Einstein
equations.  Solving the bulk equations of motion, we find
\eqn\phisol{\phi(x_5) ~=~{3\over 4} log \vert {4\over 3} x_5 + c \vert + d}
\eqn\asol{A^\prime ~=~ {1\over 3} \phi^\prime}
Notice that at $x_5 = -{3\over 4}c$, there is a singularity:
the scale factor vanishes ($A$ goes to $-\infty$), and the
$\vert {\rm curvature}\vert \to \infty$.  
If one momentarily views $x_5$ as a time-like direction, and the
slices of constant $x_5$ as 4d spatial slices in a cosmology, then
this singularity looks like a big bang or big crunch singularity
where the spatial slices collapse to zero size.

Next, we need to include the wall source terms.  For simplicity, we 
specialize to the case:
\eqn\moreconc{f(\phi) ~=~V e^{-{4\over 3}\phi}}
However, with one exception (to be mentioned later), basically all of
our considerations 
carry over for much more generic $f(\phi)$ \KSS.

From the form of the bulk solutions, it is clear that the solution with
the wall should have the general form:

\eqn\phisollef{\phi(x_5) = {3\over 4} log \vert {4\over 3}x_5 + c_1 \vert
+ d_1, ~x_5 < 0}
\eqn\phisolrig{\phi(x_5) = {3\over 4} log \vert {4 \over 4}x_5 + c_2 \vert
+ d_2, ~x_5 > 0}
and $A^\prime = {1\over 3} \phi^\prime$.

Imposing the matching conditions at the wall, we find a $Z_2$ invariant
solution (symmetric under left/right exchange) if
\eqn\bc{c_1 = -c_2 = c,~~d_1 = d_2 = d,~~e^{-{4\over 3}d} = {4\over V}
{c\over \vert c\vert}} 
with $\it arbitrary$ c.

Now, suppose we choose $c$ positive.  Then, the solution
\phisollef~\phisolrig\ has curvature singularities at $x_5 = \pm {3\over 4}c$. 
Let us suppose the space $\it ends$ at the singularities, so that
the $x_5$ dimension is effectively an interval (with the Standard Model
brane in the middle).  Then, a simple computation reveals:

\eqn\mplanck{M_4^2 \sim M_5^3 \int dx_5 ~e^{2A} ~<~\infty}
so there is indeed four dimensional gravity coupled to the brane field
theory at $x_5 = 0$.  

\subsec{Discussion of Several Important Issues} 

There are several issues that need to be addressed about this framework
for discussing the cosmological constant in wall world scenarios.

\medskip
\noindent 1)  What about bulk quantum corrections?
\medskip
\noindent In general, choosing a bulk action with vanishing 5d cosmological
term and only kinetic terms for the bulk scalar $\phi$, as in 
\fivedac, is only sensible in an approximate sense.  We are assuming
the bulk is supersymmetric, with the brane breaking supersymmetry. 
Still, eventually the interaction of bulk and brane fields will transmit
the SUSY breaking to the bulk, and there will be subleading results which
correct the action \fivedac\ and lead to slight curvature of the 
previously Poincare invariant slices.  How do we estimate the size of these
effects?

It follows from the matching conditions \bc\ that if one chooses the
brane tension $V f(\phi(0))$ to be roughly a TeV, and one fixes
$M_4 \sim 10^{19} GeV$, then the 5d Planck scale is fixed to be
$10^5 TeV$ (and the size of the $x_5$ interval is about a millimeter). 
Interactions of bulk and brane fields are suppressed by explicit powers
of ${1\over M_5}$; therefore, bulk corrections to the 4d effective
field theory will arise in a power series in $\epsilon = (TeV / M_5)$. 
Hence, in this scenario one can arrange to cancel the leading Standard
Model $(TeV)^4$ contribution to the effective 4d cosmological term,
but there will be contributions at subleading orders in $\epsilon$.
While these are too large to be tolerated given the observed value of
the cosmological constant (unless one can somehow cancel the first
few terms in the power series), they are nevertheless hierarchically
smaller than the expected answer.  So, our philosophy should be,
that we are looking for a system where the induced cosmological term
is hierarchically smaller than what is expected (and we can
postpone understanding 
how to get precisely the right magnitude of the suppression). 

\medskip
\noindent 2) We have shown there are generically Poincare invariant
solutions to the equations, independent of Standard Model parameters.
However, are there also other curved solutions, which would be
characteristic of a 4d effective field theory with nonzero cosmological
term?

\medskip
\noindent
For generic $f(\phi)$, it turns out that curved solutions with
de Sitter or anti de Sitter symmetry $\it do$ exist \KSS.  For fine tuned
$f(\phi)$, e.g. $f(\phi) \sim V e^{\pm {4\over3}\phi}$, there are
no de Sitter or anti de Sitter solutions \ADKS. 
However, in systems with massless scalar fields, the solutions which
arise when there is a nonzero cosmological term are often not
dS or AdS solutions, but instead solutions where the scalar field
is spatially varying.  A prototypical example is string theory, where
introducing a slight cosmological constant can lead to linear dilaton
solutions instead of dS or AdS \Myers.  So it seems rather likely that in
this case also, one can find solutions with 4d slices that are
characteristic of a nonzero 4d cosmological term; however a definitive
answer to this question is lacking, since the 4d effective field theory
has not been written down. 

Even given this fact, we find it quite interesting that one can find
Poincare invariant solutions as well.  It has been very hard to find
$\it any$ examples in string theory of Poincare invariant vacua without
supersymmetry.  Some candidates were proposed in \KKS, but 
have a very non-generic low energy effective field theory (with Bose-Fermi
degeneracy) and have only been studied at low orders in perturbation
theory.  
Indeed it has been advocated by Banks \Tom\ that perhaps M theory does not
admit nonsupersymmetric, Poincare invariant solutions.  We find 
it intriguing that these ``wall worlds'' do admit Poincare invariant
solutions, and are quite similar to systems one can realize in string
theory with wrapped branes.

\medskip
\noindent 3) What about the singularities?

\medskip
\noindent
Of course, the 5d effective field theory defined by \fivedac\ breaks
down in regions of large curvature.  However, it is often the case
that string theory can regulate and provide a definition of
singular geometries.  So, it is an important problem to find a 
microphysical realization of these systems, which regulates the
singularities or describes the physics occurring there.

There are obvious analogies between our $x_5$ interval and the intervals
encountered in e.g. Horava-Witten theory or Type $I^\prime$ string theory.
Instead of expanding on those here (see e.g. \KSS\ for more discussion),
we concentrate instead on the similarity to geometries arising from
RG flows in AdS/CFT duality.

Polchinski and Strassler, for instance, have studied a class of RG
flows from the deformed ${\cal N}=4$ super Yang-Mills to confining
gauge theories with less supersymmetry \PS.  Because their geometries involve
a 5d gravity theory with small curvature in the UV (near the
``Standard Model'' brane, in our language) and large curvature in the 
IR (which corresponds to our singularities), they are quite similar
to our setup.  As discussed by Bousso and Polchinski \BP,
we can then
use the results of \PS\ to infer some important ``facts'' about the singularities
we encounter here.

What Polchinski and Strassler find is that the RG flow results in
a ``discretuum'' of possible IR branes -- there are roughly
$e^{\sqrt{N}}$ possibilities for the IR brane, where $N$ is a large
number in the (super)gravity limit.  This translates immediately to the 
statement that, in our solutions,  
it is quite likely that the integration constants which arise
in $\phi(x_5)$ cannot take arbitrary values, but are
rather quantized to certain allowed values at the singularities.
The question is then, do the allowed values
allow for a solution which is closer to Poincare invariant than would 
be possible with the expected $(TeV)^4$ cosmological constant?
  
The answer seems to be yes.  As argued in \BP,\foot{This was 
independently known and stated by several others including  
N. Arkani-Hamed, E. Silverstein and the author.} 
the AdS/CFT results strongly suggest that cosmological constants which
are suppressed from $TeV$ scale by powers of $e^{-\sqrt N}$ should be
achievable.
The question of why the allowed singularity with the smallest possible
norm of the 4d cosmological constant would be cosmologically preferred
is a difficult one.  A possible scenario, following earlier work of
Brown and Teitelboim \BT, was presented in \BP (in a different context,
involving M theory compactifications with four-form fluxes).
Several aspects of their work might 
generalize to the case
under discussion.  A related approach can be found in \Frank, and a 
critical discussion appears in \MikeTom.

Another method of dealing with the issue of singularities, is to attempt
to find solutions that retain the ``self-tuning'' property (the existence of 
a flat solution independent of ``Standard Model'' parameters), but are either 
not singular or have singularities which are physically innocuous.  One
interesting approach of this sort has been detailed in \Horowitz, where
they find a self-tuning model which has 
no naked singularities and is known to arise in exact string theory constructions.
While the particular solution they find is not attractive for other reasons (there
is a strongly time-dependent 4d Newton's constant), 
the basic idea seems promising.  Another recent self-tuning 
construction which is free
of singularities appears in \Zurab.

There has been some controversy in the literature about the various
brane world approaches to studying the cosmological constant problem 
(regarding issues like physical admissibility
of the singularities).  While my point of view is well represented here,
alternative viewpoints can be found in e.g. \refs{\Witten,\Nilles}.

\newsec{Calabi-Yau Compactifications and Closed String Mirror Symmetry} 

In the next two lectures, we will work up to the study of building
blocks for microscopic ``brane worlds'' which clearly $\it are$
realizable in string theory.  
These backgrounds involve D-branes in curved geometries, and the
open string sectors which live on these branes.
Such brane theories exhibit some interesting duality
symmetries, which we will also briefly explore.
To make the discussion self-contained, we must provide a
brief description of the relevant closed string backgrounds first.

\subsec{Type II Calabi-Yau Compactifications}

Suppose one wants to find a supersymmetry-preserving compactification
of the type IIA or type IIB theory, by compactifying on a smooth manifold
$M$ of complex dimension $d$.\foot{We want to achieve Poincare
supersymmetry in the remaining $10-2d$ dimensional theory, so we will
not have to worry about e.g. the Freund-Rubin ansatz and AdS 
solutions.} 
One can argue that a necessary condition is \GSW\
\eqn\holonomy{{\rm Holonomy~ of~ M}\subset {\rm SU(d)}}

The possible choices of $M$ become more and more plentiful as $d$
is increased.  For $d=1$, the only choice is the 
two-torus $T^2$, and the resulting 8d theory has 32 supercharges.
For $d=2$, one can choose either $T^4$ or $K3$, which preserve
32 and 16 supercharges respectively.  Finally, in the case we will
utilize later on, $d=3$, there are (at least) thousands of choices
(for the earliest large compendium of such spaces that I am aware of, 
see \candelas).
The generic choice of such a complex threefold preserves 8 supercharges,
corresponding to 4d ${\cal N}=2$ supersymmetry. 
The study of such compactifications has been a rich and beautiful subject
about which we will necessarily be very brief here: for a much more 
comprehensive review, see \Greene.

These so-called $\it Calabi-Yau$ manifolds are Ricci flat and
K\"ahler.  The Ricci-flat K\"ahler metrics 
on a Calabi-Yau space $M$ come in a family of dimension
$h^{1,1}(M) + 2 h^{2,1}(M)$, where $h^{1,1}$ parametrizes the
choice of a K\"ahler form and $2 h^{2,1}$ is the dimension of
the space of inequivalent complex structures on $M$.   

A simple example of such a space is the quintic Calabi-Yau threefold in
$CP^{4}$.  $CP^4$ is defined by taking 5 homogeneous coordinates
$(z_1,\cdots,z_5)$, subject to the identification $(z_1,\cdots,z_5) \sim
(\lambda z_1,\cdots,\lambda z_5)$ where $\lambda$ is a nonzero complex
number (and with the origin deleted). 
The quintic is defined by a homogeneous equation of degree 5 in this space,
for instance
\eqn\quintic{P = \sum_{i=1}^{5} z_{i}^5~=~0} 
The complex structure deformations of this manifold are parametrized 
simply by monomial deformations of the equation \quintic, modulo linear
changes of variables $z_i \to A_{ij} z_j$.  
In the end, this leads to a 101 possible 
(complex) deformations of the equation
\quintic.
The K\"ahler deformations are, in this case, simply inherited from those
of $P^4$ -- there is a single real K\"ahler parameter, parametrizing the
overall volume.

\subsec{Spectrum of IIA or IIB String Theory on $M$}

Compactifying either type II string theory on a Calabi-Yau threefold
$M$ results in a 4d, ${\cal N}=2$ supersymmetric theory in the 
remaining $R^{3,1}$.  Such a theory admits two kinds of light
supermultiplets:

\medskip
\noindent
$\bullet$ The vector multiplet, which consists of a complex scalar field,
a vector field, and fermions, all in the adjoint representation of the
gauge group ${\cal G}.$

\medskip
\noindent
$\bullet$ The hypermultiplet, which consists of 2 complex scalars and
fermi superpartners, in any representation of the gauge group ${\cal G}$.

The moduli of the Ricci-flat metric on $M$ show up as scalars in such
multiplets in the compactified IIA/B theory.  However, the correspondence
between the geometry of $M$ and the type of multiplet is different for
the two theories. 

\noindent
{\it{Type~IIA~on~M}}:

The IIA theory in ten dimensions has a metric, an NS $B_{\mu\nu}$ field,
a dilaton $\phi$, and 1 and 3 form RR gauge fields.
On dimensional reduction on $M$, this gives rise to $h^{1,1}(M)$ 
abelian vector multiplets (where the scalars come from the real 
K\"ahler moduli of
the metric plus the $B_{\mu\nu}$ field, and the vector comes from
$C_{\mu\nu\rho}$). 
On the other hand, the complex structure moduli of the metric together
with the scalars coming from absorbing 3-forms on $M$ with $C_{\mu\nu\rho}$
give rise to (the scalar components of) $h^{2,1}(M)$ hypermultiplets.
It turns out that the dilaton in the IIA theory is also part of a hyper,
yielding a total of $h^{2,1}(M) + 1$ hypers.

\noindent
{\it{Type~IIB~on~M}}:

In the IIB theory, in addition to the metric, NS B field and dilaton,
there are 0,2 and 4 form RR gauge fields.
These give rise to $h^{2,1}(M)$ abelian 
vector multiplets in the low energy theory
(with the scalars coming from complex structure moduli, and the 
vectors coming from $C_{\mu\nu\rho\lambda}$).
On the other hand, the K\"ahler moduli, the $B_{\mu\nu}$ and
$C_{\mu\nu}$ fields (NS and RR two forms), and the RR 4-form give
rise to 
$h^{1,1}(M)$ hyper multiplets coming from the (1,1) forms on $M$.
Including the dilaton, which again transforms as part of a hyper,
this yields a total of $h^{1,1}(M)+1$ hypers.

By ${\cal N}=2$ supersymmetry, there are several simplifications in the
low energy effective action for these theories.  First of all, with this
much supersymmetry, there is no potential generated for ``flat directions''
which are present in the tree-level theory.  Hence, there are moduli spaces
${\cal M}$ of exactly degenerate supersymmetric vacua (the physical
reflection of the moduli space of Ricci-flat metrics on the Calabi-Yau
$M$, if you will).  Furthermore, because of the extended supersymmetry,
the moduli space ${\cal M}$ takes the form of a product of vector
and hypermultiplet moduli spaces:
\eqn\product{{\cal M} = {\cal M}_v \times {\cal M}_h}
where the metric on ${\cal M}_{v,h}$ is independent of 
VEVs of scalars in
the ``other'' kind of multiplet.  

\subsec{Quantum Corrections}

String theory on $M$ 
comes with two natural perturbative expansions: an expansion
in string loops, controlled by the dilaton $g_s = e^{-\phi}$, and an
expansion in sigma model perturbation theory (or curvatures), which is
roughly controlled by ${R^2\over {\alpha^\prime}}$ where $R$ is some
characteristic ``size'' of $M$ (controlled by the K\"ahler moduli). 
Large $g_s$ corresponds to strong string coupling, while large
sigma model coupling means that classical geometry is not necessarily
a good approximation, and the ``stringy'' phenomena characteristic of
quantum geometry can occur \Greene.

One can then consider corrections to the tree-level picture in both
of these expansions.  To be concrete, let's consider the geometry
(metric) of the vector multiplet moduli space ${\cal M}_v$.  It is
controlled, as is familiar from Seiberg-Witten theory \sw, by a holomorphic
prepotential $F(\phi_i)$ where $\phi_i$ are the scalar moduli in 
the vector multiplets ($F$ also determines the 
kinetic terms of the gauge fields, the so-called ``gauge coupling
functions"). 
\medskip
\noindent$\bullet$ In the IIB theory, ${\cal M}_v$ is independent of
the K\"ahler moduli and $g_s$, because both of them are in hypermultiplets
and the geometry of the vector moduli space is independent of the VEVs
of hypers.  Therefore, it is exactly determined at both string and sigma
model tree level: it is computable in terms  
of classical geometry.  To be slightly more 
precise, each Calabi-Yau manifold $M$
is characterized by a holomorphic (3,0) form $\Omega$, which is unique
up to scale.  If we let $i,j,k$ index directions in the moduli space of
complex structures, then
\eqn\prepot{{{\partial^3 F}\over{\partial\phi_i\partial\phi_j\partial\phi_k}}
\sim \int_M \Omega \wedge \partial_i \partial_j \partial_k \Omega} 
A detailed explanation of this formula can be found, for instance, in
\cdo.

\medskip
\noindent$\bullet$ In the IIA theory, there is a more complicated
story.  Because the K\"ahler moduli are in $\it vector$ multiplets now,
there $\it are$ quantum corrections to the prepotential $F$ controlled
by the sigma model coupling.  However the dilaton is still in a 
hypermultiplet, so there are no $g_s$ corrections -- $F$ is computable
at string tree level.
Considerations of holomorphy dictate that the form of $F$ is 
such that ${\partial^3}F$ will contain contributions which are
either tree-level or non-perturbative in ${R^2\over \alpha^\prime}$
(i.e. going like $e^{-{R^2\over \alpha^\prime}}$). 
This is because the K\"ahler parameter $R$ is real, and its scalar partner
(which arises from the dimensional reduction of the NS $B$ field and
complexifies it) is an axion $a$ \WW.  Although the continuous shift
symmetry for the axion can be broken non-perturbatively, 
there is a discrete symmetry under which $a$ shifts by $2\pi$ (in a 
natural normalization).  This forbids any corrections to the
prepotential in perturbation theory, but is
consistent with nonperturbative corrections. 

What is the source of the non-perturbative corrections to sigma model
perturbation theory?  At tree level in the $g_s$ expansion, the
string worldsheet is a sphere, and ``instanton'' corrections suppressed
by $e^{-R^2\over\alpha^\prime}$ can arise when the worldsheet wraps
a holomorphic sphere (of radius $R$) embedded in the Calabi-Yau 
space $M$ \DSWW.     
Denote by $H_i$ a basis for the homology 4-cycles in $M$, and
by $b_i$ a dual set of 2-forms ($i=1,\cdots,
b_{2}(M)$).
The $H_i$ are in 1-1 correspondence with the
scalars $\phi_i$ in the ${\cal N}=2$ vector multiplets. 
At large radius, when non-perturbative corrections to the sigma model
are irrelevant, there is an elegant formula expressing the prepotential
$F$ in terms of the intersection numbers of $M$ (see e.g. \cdo) 
\eqn\classical{{{\partial^3 F}\over \partial \phi_i \partial \phi_j
\partial \phi_k} \sim \int_M b_i \wedge
b_j \wedge b_k} 
As argued in \DSWW, this is corrected by instantons to an expression of
the form
\eqn\quantum{{{\partial^3 F}\over \partial \phi_i \partial \phi_j
\partial \phi_k} \sim \int_M b_i \wedge b_j \wedge b_k +
\sum_{C} \int_C b_i ~\int_C b_j ~\int_C b_k ~e^{-Area(C)\over \alpha^\prime}} 
where the sum runs over holomorphic spheres 
$C$ passing through all three of the cycles $H_{i,j,k}$. 

\subsec{Mirror Symmetry}

Mirror symmetry basically is the statement that Calabi-Yau manifolds 
naturally come
in pairs $M$ and $W$
such that the type IIA theory on $M$ is $\it exactly ~equivalent$ to the
type IIB theory on $W$ (see \Katz\ and references therein for
the development of this idea). 

A moment's thought reflects that this is an extremely nontrivial statement
about the geometry of Calabi-Yau spaces, and a powerful computational tool
for physics.  
For instance, the roles of the K\"ahler and complex structure moduli of
$M$ and $W$ will be interchanged by the symmetry, since their physical
role in the low energy effective ${\cal N}=2$ gauge theory that arises
from the string compactification is interchanged in the IIA and IIB
theories.   
For a quick
indication of the mathematical power of this statement, recall that
the intricate prepotential \quantum\ in the IIA theory will be computable
in tree-level of both $g_s$ and $\alpha^\prime$ perturbation theory in
the IIB theory, by a formula of the form \prepot.  
Equating the two makes highly nontrivial predictions about e.g.
the multiplicity of holomorphic curves in $M$; this has been exploited
to great effect beginning with the work \cdgp.

A heuristic proof of this statement (for some classes of Calabi-Yau
manifolds) was provided by Strominger, Yau and Zaslow \syz. 
Their reasoning goes roughly as follows.  Suppose the IIA theory on $M$
is really equivalent to the IIB theory on $W$.
Then the full theories, including all BPS states and their detailed
properties, should match.

Now, what are the SUSY brane configurations allowed by Calabi-Yau
geometry, that will give rise to BPS states in the two cases?
There are basically two sets of possibilities for CY threefolds \ooy:

\noindent$\bullet$  One can wrap D2, D4 or D6 branes on holomorphic
2,4 or 6 cycles.

\noindent$\bullet$ One can wrap D3 branes
on ``special Lagragian'' three-cycles;
by definition, a special Lagrangian three-cycle $\Sigma$
is a submanifold of $M$ such that the K\"ahler form $\omega$ restricts
to zero on $\Sigma$, and $Im(\Omega)\vert_\Sigma ~=~0$ as well.

Since the IIA theory only has supersymmetric Dp branes for even $p$,
while the IIB theory only has supersymmetric Dp branes for odd $p$,
the holomorphic cycles are relevant for IIA while the special Lagrangian
cycles are relevant for IIB (as long as one is focusing only on 
point particles in the transverse $R^{3,1}$).  It is a common abuse of
terminology to simply call both special Lagrangian cycles and holomorphic
cycles ``supersymmetric cycles,'' for obvious reasons. 

So, lets start by considering the simplest possible case, the D0 brane
in type IIA on $M$. The worldvolume theory is a supersymmetric quantum
mechanics with 4 supercharges, and its moduli space is intuitively just
given by the manifold $M$ itself.
Hence, if IIB on $W$ is exactly equivalent to IIA on $M$, it must contain 
a ``mirror'' supersymmetric brane whose moduli space is also $M$! 

As discussed above, it must be a D3 brane wrapping a SUSY 3-cycle
$\Sigma \subset W$.  What are the properties of $\Sigma$?  In particular,
one needs
the complex dimension of the moduli space of the wrapped D3 brane to
be 3.  By McLean's theorem \McLean, 
$\Sigma$ itself has $b_{1}(\Sigma)$ moduli as a supersymmetric cycle in
$W$.  In addition, Wilson lines of the $U(1)$ gauge field on the
wrapped D3 brane provide another $b_1(\Sigma)$ moduli.  Thus,
we learn that we must have $b_1(\Sigma) = 3$ to match the expected
dimension of the moduli space.

Furthermore, if we fix a point in the moduli space of the special
Lagrangian cycle and simply look at the Wilson lines, they give rise
to a $T^3$ factor in the moduli space of the wrapped brane.  Hence,
we learn that in some sense, $M$ must be a $T^3$ fibration!

Now obviously, switching the role of $M$ and $W$ would yield an argument
that $W$ must also be a $T^3$ fibration.  Hence, an elegant conjecture
is that both $M$ and $W$ are fibered by special Lagrangian $T^3$s,
and in particular the mirror of the D0 brane on $M$ is a D3 brane
wrapping the supersymmetric $T^3$ on $W$.   
This is intuitively sensible: T-dualizing on the 3 circles of the 
$T^3$ would turn the IIB theory into the IIA theory, and change the
D3 brane into a D0 brane. 

This chain of arguments indicates that all Calabi-Yau manifolds with
mirrors are $T^3$ fibrations; it is furthermore a constructive argument,
since one can in principle explicitly construct the mirror manifold
by T-dualizing the supersymmetric $T^3$s.  In practice this is of course
very difficult.   Indeed, simply demonstrating the supersymmetric
$T^3$ fibration is out of reach except in very special cases; weaker
results about Lagrangian fibrations do exist.  For a recent review,
see \Dave.

\subsec{An Example}

Since the previous subsection was fairly abstract, we close this
section with a (trivial) example.  Consider IIA on a $T^2$ which is a product
of two circles, $T^2 = S^1_{R_1} \times S^2_{R_2}$ where $R_{1,2}$
are the radii.  We can clearly view this as a $T^1$ (i.e., $S^1$) 
fibration over $S^1$.  
From standard results in elementary geometry, the complex structure of
this torus is parametrized by
\eqn\cstr{\tau ~\sim~i{R_2\over R_1}}
while its K\"ahler structure (or volume) goes like
\eqn\kahler{\rho \sim i R_1 R_2}
The $i$ appears in \kahler\ because the string theory modulus $\rho$
satisfies $\rho = B + iJ$, where $B$ is the NS $B$ field and $J$ is the
geometrical K\"ahler form. 

T-dualizing along the $S^1_{R_1}$ circle has the following effect.
Define
\eqn\newr{R_1^\prime ~=~{1\over R_1}} 
(we are setting the string scale to unity for simplicity in this subsection).
Then
\eqn\tnew{\tau_{new} ~=~i{R_2 \over R_1^\prime} = iR_2 R_1 = \rho_{old}}
and
\eqn\pnew{\rho_{new} ~=~i R_1^\prime R_2 = i{R_2\over R_1} = \tau_{old}}
And of course, T-dualizing along one circle exchanges the IIA and IIB
theories.

We have succeeded in taking IIA on a torus with 
(complex,K\"ahler) moduli $(\tau,\rho)$ to
IIB on a torus with moduli $(\rho,\tau)$. 
This is mirror symmetry for $T^2$, and it has precisely arisen here
as T-duality on the $S^1$ ``fibration.''  
One can do the slightly less trivial case of $K3$ with as much
success, by viewing $K3$ as a $T^2$ fibration \syz.

\newsec{Open Strings and Mirror Symmetry}

In the previous lecture, we saw that Calabi-Yau threefolds come in
pairs $M$, $W$ such that the IIA theory on $M$ is equivalent to the
IIB theory on $W$.  This yields a powerful tool for the study of
4d ${\cal N}=2$ supersymmetric string vacua. 

By a slightly more elaborate construction, we can also manufacture
4d ${\cal N}=1$ models starting 
with type II strings on Calabi-Yau spaces.
Namely, we should compactify the type II theory on a Calabi-Yau,
and then introduce additional (space-filling) $D(p+3)$ branes
wrapping supersymmetric $p$ cycles.\foot{In the full construction,
one will also have to introduce orientifolds to
cancel the RR tadpoles.} 
It is natural to ask: what does mirror symmetry do for us in this
context? 

Let's begin the discussion in type IIA string theory.  If we wish to make
a ``brane world'' in type IIA string theory by 
compactifying on a Calabi-Yau $M$ and then wrapping $D(p+3)$ branes on
$p$ cycles in $M$, and we also want to preserve 4d ${\cal N}=1$ supersymmetry,
then the only possibility is to wrap $D6$ brane(s) on supersymmetric
(special Lagrangian) three-cycles.
Recall that a 3-cycle $\Sigma \subset M$ is called special Lagrangian iff 

\noindent$\bullet$ $\omega \vert_{\Sigma} ~=~0$ 

\noindent$\bullet$ $Im(\Omega)\vert_{\Sigma} ~=~0$ 

\noindent
where $\omega$ is the K\"ahler form of $M$, and $\Omega$ is the holomorphic
(3,0) form.
Such cycles are volume minimizing in their homology class.

\subsec{How to produce examples of $\Sigma$}

Although quite generally it is difficult to produce examples of special
Lagrangian 3-cycles in compact Calabi-Yau manifolds, there is a rather
special construction that can be used to give a simple class of examples.
Suppose we have local complex coordinates $z_{1,2,3}$ on $M$, chosen so
that: 
\eqn\kform{\omega \sim \sum_i dz_i \wedge d\overline z_i}
\eqn\threef{\Omega \sim dz_1 \wedge dz_2 \wedge dz_3}
Furthermore, 
suppose that $M$ comes equipped with a so-called ${\it real~involution}$
${\cal I}$, 
which acts at
\eqn\invol{{\cal I}:~z_i \to \overline z_i}

Consider now the fixed point locus of ${\cal I}$ in $M$, i.e. the locus of
points where $z_i = \overline z_i$.  Let us call this $\Sigma_{\cal I}$.  
It is clear from \kform\ and \threef\ that ${\cal I}$
acts on the K\"ahler form and the holomorphic three-form as
\eqn\iac{{\cal I}:~\omega \to -\omega,~~\Omega \to {\overline \Omega}}
So in particular, we read off from \iac\ that 
on $\Sigma_{\cal I}$:
\eqn\fplocus{\omega\vert_{\Sigma_{\cal I}} = 0,~~Im(\Omega)\vert_{\Sigma_{\cal I}}
= 0}
Hence, the fixed point locus $\Sigma_{\cal I}$ of a real involution ${\cal I}$
acting on $M$ is always a special Lagrangian cycle. 

Let's be very concrete by working out an example.  Consider the Calabi-Yau
hypersurface in $WP^{4}_{1,1,2,2,2}$ defined by the equation: 
\eqn\cyeq{p = z_1^8 + z_2^8 + z_3^4 + z_4^4 + z_5^4 - 2(1+\epsilon) z_1^4 z_2^4
~=~0}
Notice that $p=dp=0$ is soluble when $\epsilon \to 0$, indicating that the
hypersurface \cyeq\ becomes singular at that point in moduli space.
We will consider the region of small positive $\epsilon$.

Now, consider the real involution ${\cal I}: z_i \to \overline z_i$ acting on
\cyeq.  
The fixed point locus is obviously the locus where all of the $z_i$ are real.
What is its topology?
Let us define $u = z_1^4$, and work (without loss of generality) in the
$z_2 = 1$ patch.  Then $p=0$ implies
\eqn\quadratic{u^{2} - 2(1+ \epsilon) u + 1 + Q ~=~0}  
where
\eqn\qdef{Q \equiv z_3^4 + z_4^4 + z_5^4}
Solving \quadratic\ we find
\eqn\solquad{u_{\pm} ~=~1+\epsilon \pm \sqrt{\epsilon^2 + 2\epsilon - Q}}

What is the point of this?  The solutions \solquad\ go imaginary
for large $Q$, so $Q$ is bounded to lie in some domain of size basically
$2\epsilon$ (for small $\epsilon$). 
The locus $Q < 2\epsilon$ intersects the fixed point locus of
${\cal I}$ in a 3-ball $B_3$, and has boundary $Q = 2\epsilon$ which is 
(up to finite covering) an $S^2$.  The two different branches of solutions for
$u$ in \solquad\ are glued together along this boundary $S^2$; so altogether
$\Sigma_{\cal I}$ consists of two $B_3$s glued together on an $S^2$.  
But of course this is nothing but an $S^3$.  

It follows from these manipulations that the size of the $S^3$ goes to zero
as $\epsilon \to 0$; the singular point in moduli space is related to 
the existence of this collapsing 3-cycle. 

\subsec{D6 Branes wrapping special Lagrangian cycles}

Now that we have gotten some feeling for very simple examples of special
Lagrangian cycles, lets start to consider the physical theory living on a
$D6$ brane which wraps such a cycle $\Sigma$.  
Since the brane breaks half of the supersymmetry,
the low energy theory on the brane (living in the noncompact $R^{3,1}$) will
be a 4d ${\cal N}=1$ field theory.
The $D6$ brane gauge field will descend to yield a $U(1)$ gauge supermultiplet in
4d.  The other kind of light multiplet in ${\cal N}=1$ theories is the chiral
multiplet; how many of these will be present?

It follows from the work of McLean \McLean\ that the ``geometrical'' moduli
space of $\Sigma$  
has (unobstructed) real dimension $b_{1}(\Sigma)$.  String theory complexifies
this with Wilson lines of the $U(1)$ gauge field, yielding a 
moduli space of vacua with $b_1(\Sigma)$ complex dimensions for the brane
worldvolume field theory. 

What are good coordinates on this moduli space?
Choose a basis $\gamma_j$ for $H_1(\Sigma)$, and choose discs $D_j \subset M$
such that $\partial D_j = \gamma_j$.  Define
\eqn\omegajdef{\omega_j = \int_{D_j} \omega}  
which is the area that a holomorphic disc in $D_j$s relative
homology class would
have (if it existed). 
To complexify this, consider also the $b_1(\Sigma)$ Wilson lines
\eqn\wilsdef{a_j ~=~\int_{\gamma_j} A}
where $A$ is the $U(1)$ gauge field on the $D6$ brane. 
Together, \omegajdef\ and \wilsdef\ yield the scalar components of 
$b_1(\Sigma)$ chiral multiplets $\phi_j$ which live in the 4d ${\cal N}=1$ theory
on the wrapped brane \refs{\Vafamir,\kklmI}:
\eqn\chiral{\phi_j ~=~{\omega_j \over \alpha^\prime} + i a_j + \cdots} 

Now, in any ${\cal N}=1$ supersymmetric field theory, a quantity of great interest
which governs the vacuum structure and tends to be exactly computable is the
superpotential $W(\phi)$ as a function of the chiral fields $\phi$.  
How do we compute $W(\phi_j)$ in the theories at hand?  

First of all, there is no superpotential for the $\phi_j$ to all orders in
$\alpha^\prime$.  The proof of this statement is quite analogous to the 
one used in discussions of heterotic string compactifications \DSWW.  Because
the Wilson line $a_j$ has a shift symmetry under large gauge transformations,
$W(\phi_j)$ must not have any polynomial dependence on $a_j$.  But since
$a_j$ appears in the chiral field $\phi_j$ as in \chiral, and $W$ is a 
holomorphic function of the chiral fields, this implies that there can be
no polynomial terms in $W(\phi_j)$ at all. 
This is consistent with McLean's result in pure mathematics, which roughly
speaking sees $\alpha^\prime$ perturbation theory.
 
On the other hand, terms of the form
\eqn\inst{e^{-(w_j + ia_j)} ~=~e^{-\phi_j}}
are consistent with shifts $a_j \to a_j + 2\pi$ which occur under large
gauge transformations, and hence such terms in the
superpotential cannot be ruled out.
What would the source of such terms be?  Just as closed string theories have
worldsheet instantons, D brane theories on Calabi-Yau spaces can have
``disc instantons.''  
At tree level, the open string worldsheet is a disc $D$.  One can consider 
holomorphic maps $D \to M$ such that $\partial D = \gamma_j \subset \Sigma$,
and such that the normal derivative to the map at the boundary is in the
pullback of the normal bundle to $\Sigma$ in the Calabi-Yau.
The claim is then that the superpotential in these $D6$ brane theories is
entirely generated by such disc instanton effects.

For instance, one can formally compute couplings like the $F_i \phi_j \phi_k$
coupling that would arise between two scalars and the auxiliary field $F$
in chiral multiplets if there is a nontrivial superpotential.
This is discussed at length in \kklmI\ (and is very closely related to the
discussion in \Vafamir).  
The upshot is that one can give a formula for this three-point function on
the string worldsheet in terms of an infinite sum over disc instantons.
If we call the vertex operators for the spacetime fields appearing in
this coupling $V^i_F, V^j_\phi, V^k_\phi$, then the three-point function
$\langle V^i_F V^j_\phi V^k_\phi \rangle$ has the following expression. 
Denote by $d^{\{n_a\}}_{\{m_l\}}(i,j,k)$ the number of holomorphic maps from the
disc $D$ to $M$ with the following properties:

\noindent
i) $[\partial D] ~=~\sum_l m_l \gamma_l$

\noindent
ii) $V^{i,j,k}$ are mapped in cyclic order to the intersection of $\partial D$
with the 2-cycles in $\Sigma$ dual to $\gamma_{i,j,k}$.

\noindent
iii) $[\partial D - \sum_l m_l D_l]$, which by i) is a closed 2-cycle in $M$,
satisfies
\eqn\twocycle{[\partial D - \sum_l m_l D_l] ~=~\sum_a n_a K_a} 
where the $K_a$ are a basis for $H_2(M)$.   

Then one can derive the statement:

\eqn\threept{\langle V^i_F V^j_\phi V^k_\phi \rangle \sim 
\sum_{m_l, n_a \geq 0} ~\int_{\partial D} \theta^i \int_{\partial D}\theta^j
\int_{\partial D} \theta^k~d^{\{n_a\}}_{\{m_l\}}(i,j,k) 
~\prod_{l=1}^{b_1(\Sigma)} e^{-m_l \phi_l}~\prod_{a=1}^{h^{1,1}(M)} e^{-n_a t_a}}
where $\theta^i$ is the harmonic one-form associated to $\gamma_i$, and
$t^a = \int_{K_a} \omega$ is the area of $K_a$.
This formula is, in some natural sense, the open string analogue of the
instanton sum formula 
\quantum\ for the prepotential in closed string Calabi-Yau compactifications.
In some very special examples, such disc instanton sums have proven
directly computable \OgVafa.

\medskip
\noindent{\it Closed String Parameters}

How do the closed string moduli of the Calabi-Yau $M$ play a role in the
brane theory?  
From \threept\ above, it is clear that the superpotential $W$ of the brane
theory really depends on the closed string K\"ahler moduli; they enter as
$\it parameters$ $t_a$, so we should really denote $W$ as $W(\phi_j;t_a)$ to
indicate the relevance of the closed string background.   
In fact, it was argued rather generally in \bdlr\ 
that in this class of theories (so-called ``A-type'' branes), 
the K\"ahler (complex structure) moduli of $M$ will only
enter in the superpotential (FI D-terms) of the wrapped brane theory.
This is in accord with \threept, where the K\"ahler dependence is manifest
and there is no explicit complex structure dependence.
The dependence of the FI D-terms on the complex structure of $M$ has been
explored, for instance, in \refs{\McGreevy,\Douglas}.

\subsec{Type IIB ``Mirror'' Brane Worlds}

Thus far, we have been focusing our attention on the brane worlds we can
construct in the IIA theory, but of course it is possible to make analogous
type IIB constructions by wrapping 5,7 or 9 branes on holomorphic 2,4 or 6
cycles (or indeed, by having D3 branes transverse to the Calabi-Yau).

In \S4.4, we saw that mirror symmetry was of great use in ``solving'' the
${\cal N}=2$ theories that come out of string theory, by making the
prepotential, which receives an infinite series of quantum corrections
in the IIA picture \quantum, explicitly computable at tree level in the
IIB picture via \prepot.  Mirror symmetry should
be a similarly powerful tool in studying brane worlds of the type under
discussion.  Computations of superpotentials, which by \threept\ are 
dauntingly difficult in the IIA picture, are much simpler in the  
IIB picture.

In fact, it was argued in \bdlr\ that in the case of $D(p+3)$ branes wrapping
holomorphic $p$ cycles, the superpotential is exact at tree level (receives
no $\alpha^\prime$ corrections whatsoever).  Hence, in the IIB theory,
superpotentials are effectively as computable as e.g. prepotentials in
the closed string case.   
The challenge, then, is to find a mirror IIB brane configuration for a 
IIA configuration consisting of a $D6$ brane wrapping a special Lagrangian
three-cycle $\Sigma \subset M$.  Clearly, the IIB theory will be compactified
on the mirror manifold $W$; the question is, what is the mirror brane setup?

To find concrete examples it is most convenient to focus on cases where
the mirror IIB setup turns out to be a $D5$ brane wrapping a rational curve
$C \subset W$.   
The generalities of this kind of correspondence were discussed in 
\kklmI\ and very concrete examples, where a disc instanton generated superpotential
in the IIA theory maps to a tree level superpotential in the IIB theory, were
presented in \kklmII.

So, what is the physics of a IIB $D5$ brane wrapping $C$?
As always, there is a $U(1)$ gauge supermultiplet.  The number of massless
chiral multiplets is given by the number of small 
deformations of the curve,
parametrized by $H^{0}(C,N_C)$, where $N_C$ is the normal bundle of 
$C \subset M$.\foot{If one were to wrap a higher genus curve, there would
of course also be Wilson lines; but flat bundles on $P^1$ do not lead to
any additional degrees of freedom.} 
However, by classical deformation theory, deformations of $C \subset W$
can be ${\it obstructed}$; this corresponds precisely to a massless chiral
multiplet which has a nontrivial higher order superpotential! 
If $h^{0}(C,N_C)=1$, and we call the chiral
multiplet $\tilde \phi$, then an $N$th order obstruction is reflected in 
a superpotential on the brane which looks like
\eqn\obst{\tilde W \sim \tilde \phi^{N+2}} 

As in the IIA $D6$ brane theory, ${\it closed}$ string moduli enter in the IIB
$D5$ brane theory as parameters in the Lagrangian.  From \bdlr, the
superpotential depends only on the complex structure of $W$, while the
FI D-terms for the $U(1)$ gauge field can depend on the K\"ahler structure of 
$W$.  In concrete examples, the moduli space of $C$ can exhibit very
intricate behavior as one varies the complex structure parameters
$\psi_a$ of $W$.  So we should write the superpotential as 
$\tilde W(\tilde \phi_i;\psi_a)$
where $a$ runs over the complex structure deformations of $W$, and 
the $\tilde \phi_i$ parametrize the deformations of $C$.

Therefore, if one finds a mirror pair consisting of a $D6$ brane wrapping
a special Lagrangian cycle $\Sigma \subset M$ and a $D5$ brane wrapping
$C \subset W$, 
then the IIA disc instanton sum \threept\ should basically map to
${\it purely~ classical}$ geometrical data
in the IIB theory. The superpotential $W(\phi_i;t_a)$ in IIA will encode
the same data as $\tilde W(\tilde \phi_i;\psi_a)$.  The map between the parameters
$t_a$ and $\psi_a$ will of course be the mirror map between the closed
string moduli spaces.  On the other hand, working out the map between
the $\it open$ string fields $\phi$ and $\tilde \phi$ is a complicated
problem about which little is known at this point \kklmII.  

In practice, how does one go about constructing such mirror pairs?
The strategy followed in \kklmII\ was to wrap $D5$ branes on  
curves $C \subset W$ which collapse to zero volume at some particular
point in the K\"ahler moduli space of $W$.  Then the 3-cycle $\Sigma$
that the mirror $D6$ brane wraps must collapse at the mirror point
in the complex structure moduli space of $M$.\foot{A simple argument for
this is that in the closed string theory, there will be light non-perturbative
states associated with the collapsing curve; to replicate this phenomenon,
there must also be a vanishing cycle on the mirror manifold.}    
In examples where $b_1(\Sigma) > 0$ but the curve $C$ has less than
$b_1(\Sigma)$ unobstructed deformations, the $D6$ theory must
``lose'' some its tree-level moduli to an instanton generated potential.
Examples of this sort were produced in \kklmII, which is strong evidence
for the presence of the disc instanton effects \threept.
It would be extremely interesting to actually find an open string
analogue of the mirror map, which lets one directly map the IIA
superpotential to the IIB superpotential.  As a byproduct, one might
obtain nice counting formulas for holomorphic discs with boundaries
in a special Lagrangian cycle \Vafamir.

\bigskip
\centerline{\bf{Acknowledgements}}

These lecture notes are being submitted to the proceedings of both
TASI 1999: Strings, Branes and Gravity, and the 2000 Trieste Spring
Workshop on Superstrings and Related Matters.
I would like to thank the local organizers of TASI 1999 for providing
such a wonderful atmosphere for the school, and the students for
their enthusiastic participation. 
I would also like to thank the organizers of the 2000 Trieste Spring Workshop
on Superstrings and Related Matters for
providing a very stimulating environment in which to deliver these
lectures.
My thinking about the subjects in the first two lectures was developed
in collaborations with M. Schulz and E. Silverstein, while for the latter
two it was developed in collaborations with 
S. Katz, A. Lawrence and J. McGreevy.
In addition, I would like to acknowledge discussions with 
T. Banks, S. Dimopoulos, N. Kaloper,
J. Maldacena, S. Shenker, R. Sundrum, L. Susskind, S. Thomas,
H.Verlinde and E. Witten 
which greatly influenced my thinking on some of these subjects.
Some of the research described in these lectures occurred while
the author was enjoying the hospitality of the Aspen Center for Physics
and the Institute for Advanced Study in Princeton. 
This work was supported in part by 
an A.P. Sloan Foundation Fellowship, a DOE OJI Award, and the DOE under
contract  
DE-AC03-76SF00515.

\listrefs
\end